\documentclass[aps, prl, reprint, floatfix,
  preprintnumbers, showpacs, showkeys]{revtex4-1}
\usepackage{dcolumn,graphicx}
\usepackage{mathptmx}
\usepackage[colorlinks=true, linkcolor=blue, citecolor=blue,
filecolor=blue, urlcolor=blue]{hyperref}

\newcommand{\q}[2]{\ensuremath{#1\ \mathrm{#2}}}

\newcommand{\code}[1]{\texttt{\small #1}}

\newcommand{\Ie}{\ensuremath{I}}
\newcommand{\ri}{\ensuremath{r}}

\newcommand{\irel}{\ensuremath{n}}
\newcommand{\Ndot}{\ensuremath{\dot{n}}}
\newcommand{\Na}{\ensuremath{N_a}}
\newcommand{\Nc}{\ensuremath{N_c}}

\newcommand{\lrel}{\ensuremath{\ell}}
\newcommand{\Ldot}{\ensuremath{\dot{\ell}}}
\newcommand{\La}{\ensuremath{L_a}}
\newcommand{\Lc}{\ensuremath{L_c}}

\begin{document}

\newlength{\figwid}
\setlength{\figwid}{0.8\columnwidth}

\date{\today}

\title{Collimation with hollow electron beams}

\author{G.~Stancari}
\thanks{Corresponding author}
\email[e-mail: ]{stancari@fnal.gov}
\altaffiliation{on leave from Istituto
    Nazionale di Fisica Nucleare (INFN), Sezione di Ferrara, Italy.}
\author{A.~Valishev}
\author{G.~Annala}
\author{G.~Kuznetsov}
\thanks{Deceased.}
\author{V.~Shiltsev}
\author{D.~A.~Still}
\author{L.~G.~Vorobiev}
\affiliation{Fermi National Accelerator Laboratory, P.O. Box 500,
  Batavia, IL 60510, U.S.A.}

\begin{abstract}
  A novel concept of controlled halo removal for intense high-energy
  beams in storage rings and colliders is presented. It is based on
  the interaction of the circulating beam with a 5-keV, magnetically
  confined, pulsed hollow electron beam in a 2-m-long section of the
  ring. The electrons enclose the circulating beam, kicking halo
  particles transversely and leaving the beam core unperturbed. By
  acting as a tunable diffusion enhancer and not as a hard aperture
  limitation, the hollow electron beam collimator extends conventional
  collimation systems beyond the intensity limits imposed by tolerable
  losses. The concept was tested experimentally at the Fermilab
  Tevatron proton-antiproton collider. The first results on the
  collimation of 980-GeV antiprotons are presented.
\end{abstract}



\preprint{FERMILAB-PUB-11-192-AD-APC}

\maketitle

In high-energy particle accelerators and storage rings, the
collimation system must protect equipment from intentional and
accidental beam aborts by intercepting particle
losses~\cite{Tev_coll,LHC_coll,Seidel:1994}. Its functions include
controlling and reducing the beam halo, which is continually
replenished by various processes such as beam-gas scattering,
intrabeam scattering, electrical noise in the accelerating cavities,
ground motion, betatron resonances, and beam-beam
collisions. Uncontrolled losses of even a small fraction of the
circulating beam can damage components, quench superconducting
magnets, or produce intolerable experimental backgrounds. Collimators
also serve as a diagnostic tool for fundamental machine measurements,
such as transverse admittances, beam vibrations, and diffusion rates.

Conventional collimation schemes are based on scatterers and
absorbers, possibly incorporating several stages. The primary
collimators (or targets) are the devices closest to the beam. They
generate random transverse kicks mainly via multiple Coulomb
scattering. In the Tevatron, the primary collimators are 5-mm tungsten
plates positioned about 5~standard deviations~($\sigma$) away from the
beam axis. The random multiple-scattering kick has a root mean square
(r.m.s.) of \q{17}{\mu rad} for 980-GeV protons. The betatron
oscillation amplitude of the affected particles increases, and a large
fraction of them is captured by the secondary collimators (or
absorbers), suitably placed around the ring. In the Tevatron, the
absorbers are 1.5-m steel blocks at 6$\sigma$.

The conventional two-stage system offers robust shielding of sensitive
components and it is very efficient in reducing beam-related
backgrounds at the experiments. However, it has limitations. In
high-power accelerators, the minimum distance between the collimator
and the beam axis is limited by instantaneous loss rates, radiation
damage, and by the electromagnetic impedance of the device. Moreover,
beam jitter, caused by ground motion and other vibrations and partly
mitigated by active orbit feedback, can cause periodic bursts of
losses at aperture restrictions.

\begin{figure}
\includegraphics[width=0.7\columnwidth]{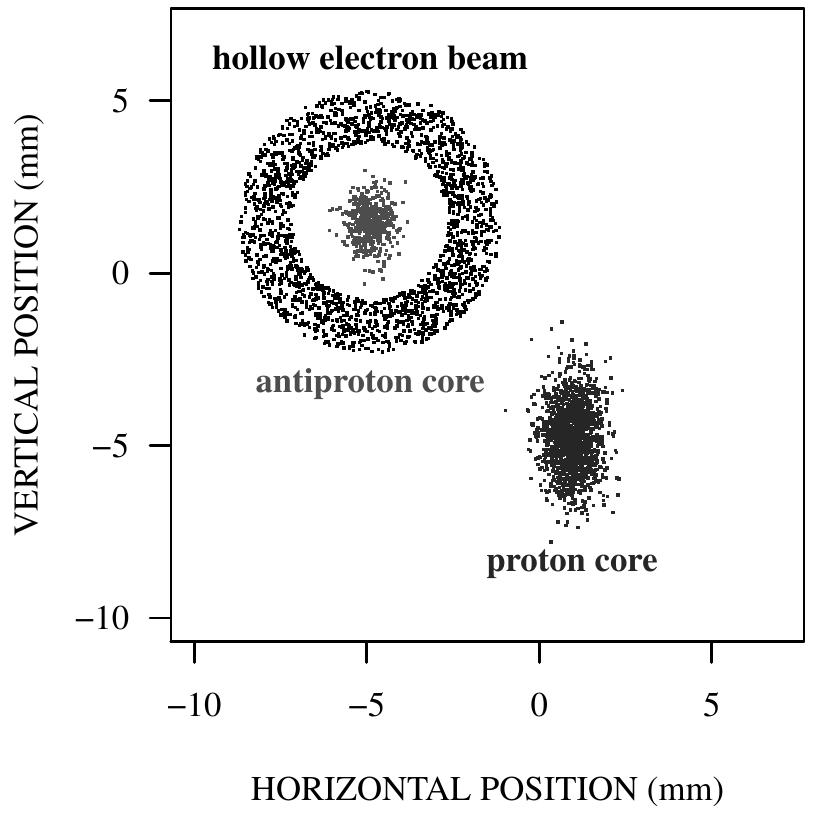}\\
\includegraphics[width=\columnwidth]{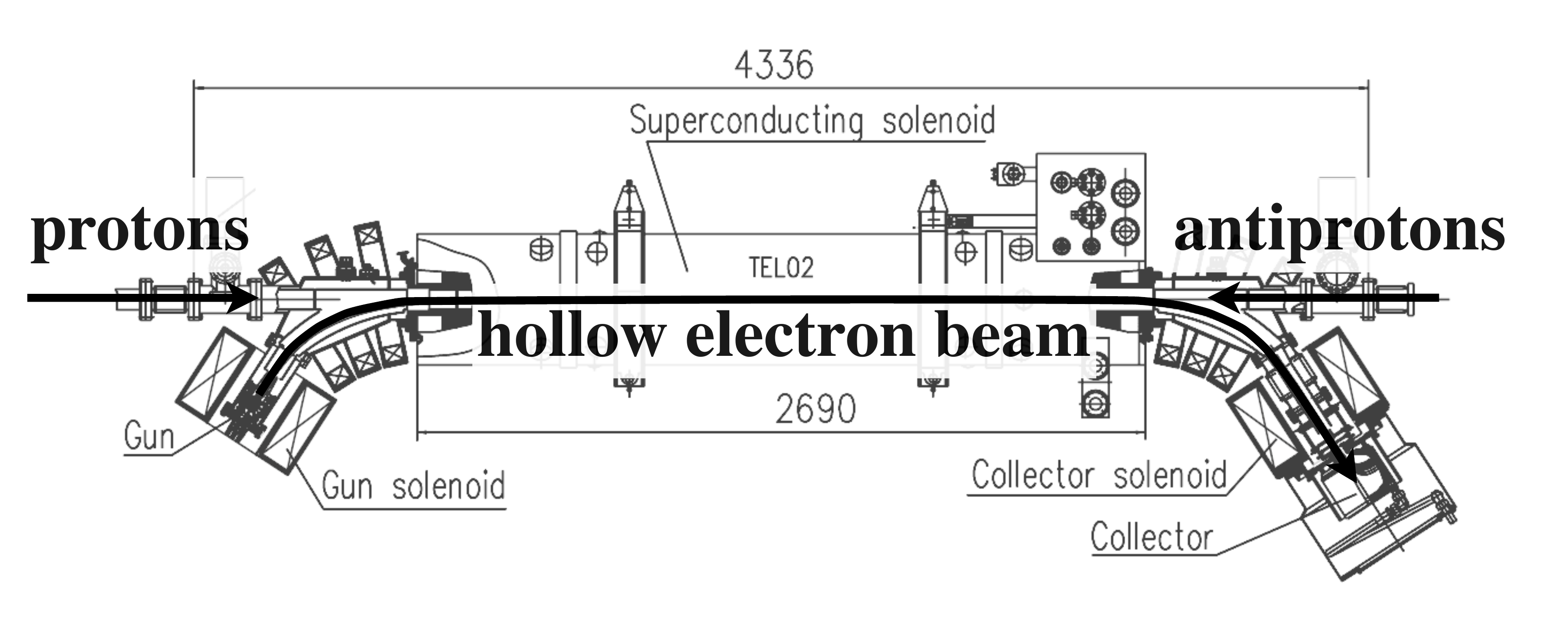}
\caption{Layout of the beams in the Tevatron.}
\label{fig:apparatus}
\end{figure}

\begin{figure}
\begin{tabular}{lr}
(a) & (b) \\
\includegraphics[width=0.53\columnwidth]{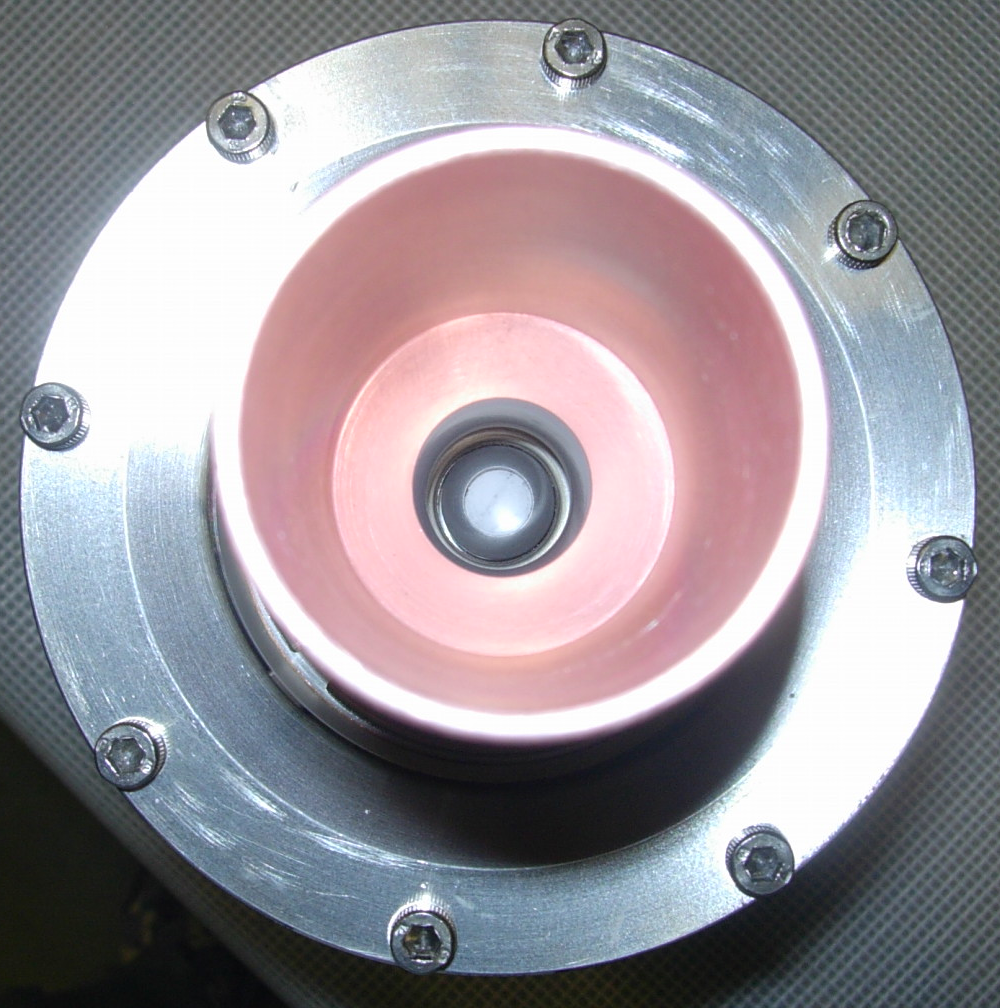} &
\includegraphics[width=0.43\columnwidth]{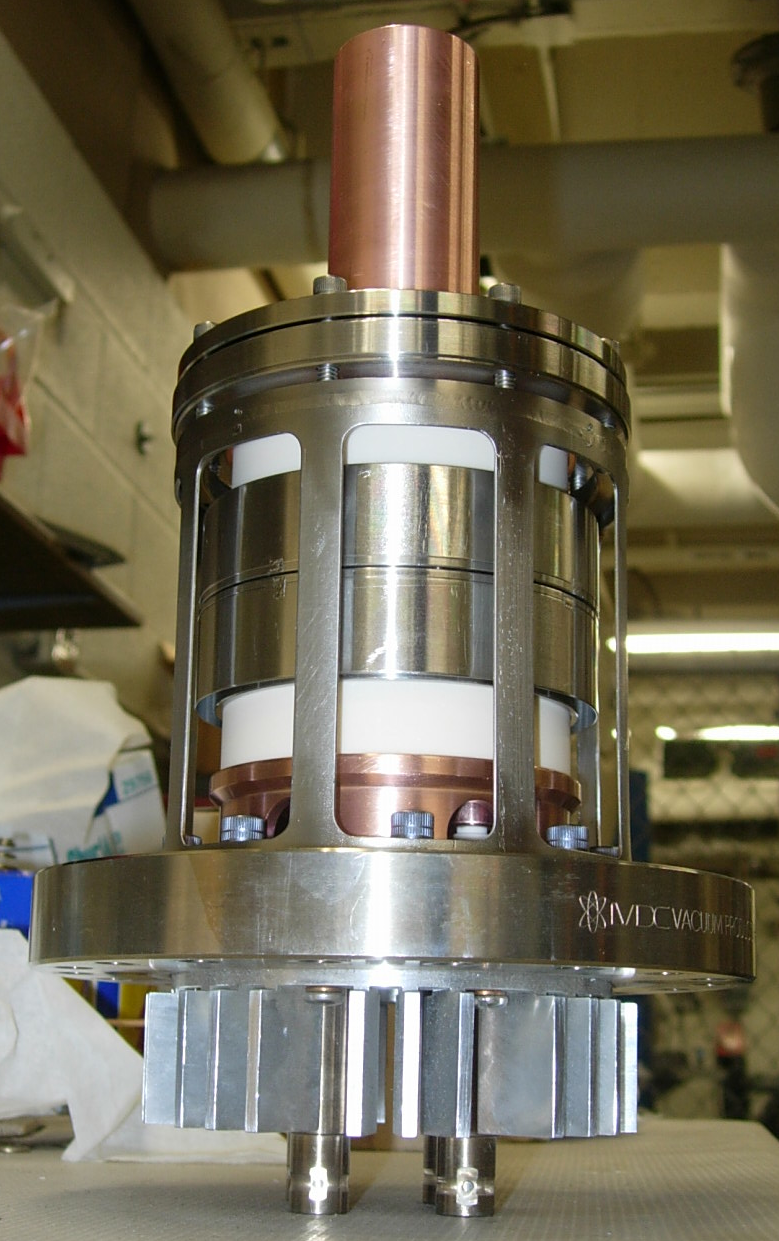} \\
\includegraphics[width=0.53\columnwidth]{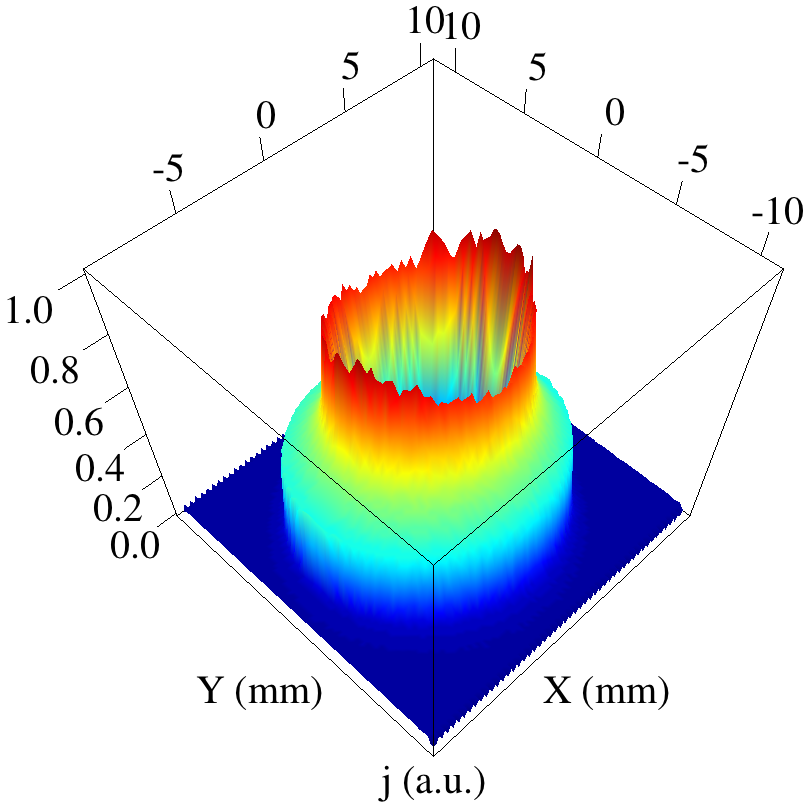} &
\includegraphics[width=0.43\columnwidth]{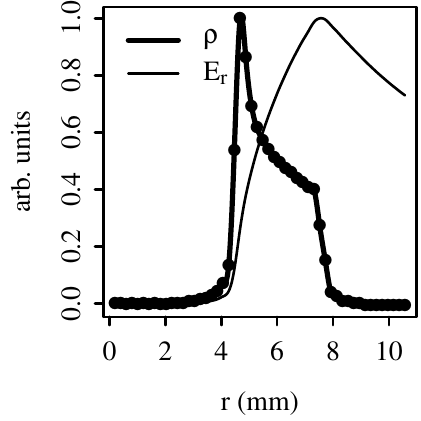} \\
(c) & (d) \\
\end{tabular}
\caption{(Color online.) Hollow electron gun: (a)~top view; (b)~side
  view; (c)~measured current density profile; (d)~measured charge
  density~$\rho$ and calculated radial electric field~$E_r$.}
\label{fig:egun}
\end{figure}

The object of this research is whether the hollow electron beam
collimator (HEBC) is a viable complement to conventional systems in
high-intensity storage rings and colliders, such as the Tevatron or
the LHC~\cite{HEBC_concept, Shiltsev:EPAC:2008, Smith:PAC:2009,
  HEBC_previous}. In a hollow electron beam collimator, electrons
enclose the circulating beam over a 2-m section of the ring immersed
in a 1~T to 3~T solenoidal field (Figure~\ref{fig:apparatus}). The
electron beam is generated by a pulsed 5-kV electron gun and it is
transported with strong axial magnetic fields, in an arrangement
similar to electron cooling~\cite{Parkhomchuk:RAST:2008} and electron
lenses~\cite{TEL}. Its size in the interaction region is controlled by
varying the ratio between the magnetic fields in the main solenoid and
in the gun solenoid. Halo particles experience nonlinear transverse
kicks and are driven towards the collimators. If the hollow current
distribution is axially symmetric there are no electric or magnetic
fields inside and the beam core is unperturbed.  A magnetically
confined electron beam is stiff, and experiments with electron lenses
show that it can be placed very close to, and even overlap with the
circulating beam. Another advantage is that, contrary to conventional
systems, no nuclear breakup is generated in the case of ion
collimation.

The transverse kick~$\theta$ experienced by particles of magnetic
rigidity~$(B\rho)_p$ traversing a hollow electron beam at a
distance~$r$ from its axis depends on the enclosed electron
current~$I_r$ and on the length~$L$ of the interaction region:
\begin{equation}
  \theta = \frac{2 \, I_r \, L \, (1\pm \beta_e
    \beta_p)}{r \, \beta_e \, \beta_p \, c^2 \, (B\rho)_p} \left(\frac{1}{4\pi
\epsilon_0}\right),
\end{equation}
where $\beta_e c$ is the electron velocity and $\beta_p c$ is the
particle velocity. The $+$~sign applies when the magnetic and electric
forces have the same direction.  For example, in a setup similar to
that of the Tevatron electron lenses ($I_r=\q{1}{A}$, $L=\q{2}{m}$,
$\beta_e = 0.14$, $r=\q{3}{mm}$), the corresponding radial kick is
\q{0.3}{\mu rad} for 980-GeV counterpropagating antiprotons.  The
intensity of the transverse kicks is small and tunable: the device
acts more like a soft scraper or a diffusion enhancer, rather than a
hard aperture limitation. Because the kicks are not random in space or
time, resonant excitation is possible if faster removal is desired.

Analytical expressions for the current distribution were used to
estimate the effectiveness of the HEBC on a proton beam. They were
included in tracking codes such as \code{STRUCT}, \code{LIFETRAC}, and
\code{SixTrack}~\cite{tracking_codes} to follow core and halo
particles as they propagate in the machine lattice. These codes are
complementary in their treatment of apertures, field nonlinearities,
and beam-beam interactions. Preliminary simulations suggested that
effects would be observable and that measurements would be compatible
with normal collider operations.

The concept was tested experimentally in the Fermilab Tevatron
collider. In the Tevatron, 36 proton bunches collide with 36
antiproton bunches at an energy of 980~GeV per beam. Each particle
species is arranged in 3~trains of 12~bunches each. Initial beam
intensities are typically \q{3\times 10^{11}}{protons/bunch} and
\q{10^{11}}{antiprotons/bunch}. Beam lifetimes range between 10~h and
100~h. There are 2 head-on interaction points, corresponding to the
CDF and the DZero experiments. The maximum luminosity is \q{4\times
  10^{32}}{cm^{-2} \, s^{-1}}.  The machine operates with betatron tunes
near~20.58.

A 15-mm-diameter hollow electron gun was designed and built
(Figure~\ref{fig:egun}). It is based on a tungsten dispenser cathode
with a 9-mm-diameter hole bored through the axis of its convex
surface. The peak current delivered by this gun is 1.1~A at 5~kV. The
current density profile was measured on a test stand by recording the
current through a pinhole in the collector while changing the position
of the beam in small steps. A sample measurement is shown in
Figure~\ref{fig:egun}. The gun was installed in one of the Tevatron
electron lenses, where the pulsed electron beam could be synchronized
with practically any bunch or group of bunches.

\begin{figure}
\centering
\includegraphics[width=\figwid]{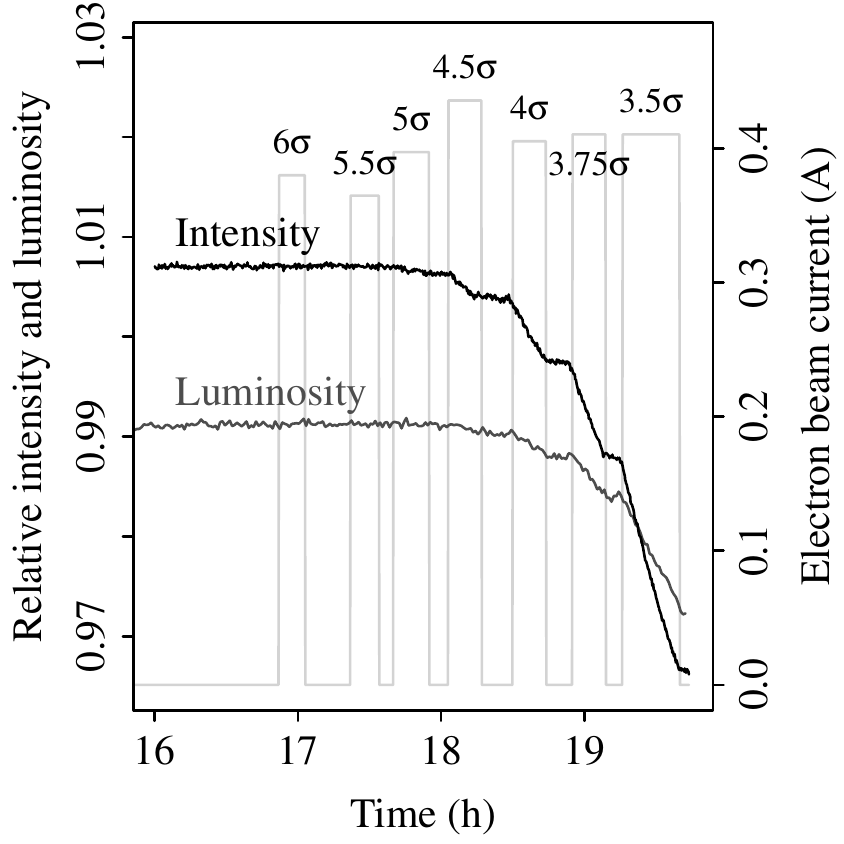}
\caption{Relative intensity and luminosity of the affected bunch
  train, for different transverse sizes of the electron beam. The
  light-gray trace is the electron beam current (right axis).}
\label{fig:removal}
\end{figure}

\begin{table}
\begin{ruledtabular}
\begin{tabular}{rdddd}
  \multicolumn{1}{c}{\Ie} & \multicolumn{1}{c}{\ri} &
  \multicolumn{1}{c}{\Ndot} & \multicolumn{1}{c}{\Ldot} &
  \multicolumn{1}{c}{$\Ldot/\Ndot$} \\
  \multicolumn{1}{c}{mA} & \multicolumn{1}{c}{$\sigma_y$} &
\multicolumn{1}{c}{\%/h} & \multicolumn{1}{c}{\%/h} & \\
  \hline
  0 &      &  0.009(5) & 0.03(1) & \\
  380 & 6.0  &  0.03(5) & 0.3(2) & 9(7) \\
  366 & 5.5  & -0.07(4) & -0.09(9) & 1(1) \\
  397 & 5.0  & -0.31(3) & -0.06(9) & 0.2(3) \\
  436 & 4.5  & -1.32(4) & -0.5(1) & 0.34(7) \\
  405 & 4.0  & -2.49(3) & -0.78(9) & 0.32(4) \\
  410 & 3.75 & -3.83(3) & -1.83(9) & 0.48(2) \\
  410 & 3.5  & -5.18(2) & -2.65(4) & 0.512(7) \\
\end{tabular}
\end{ruledtabular}
\caption{Relative particle removal rates~\Ndot\ and luminosity decay
  rates~\Ldot\ as a function of total electron beam current~\Ie\ and
  hole radius~\ri.}
\label{tab:rates}
\end{table}

The behavior of the device and the response of the circulating beams
were measured for different beam currents, relative alignments, hole
sizes, pulsing patterns, and collimator system configurations. Here,
we focus on a few representative experiments illustrating the main
effects of the electron beam acting on antiproton bunches.  Other
important effects, such as collimation efficiencies, fluctuations in
losses, and diffusion rates will be presented in a separate
report. Antiprotons were chosen for two main reasons: their smaller
transverse emittances (achieved by stochastic and electron cooling)
made it possible to probe a wider range of confining fields and hole
sizes; and the betatron phase advance between the electron lens and
the absorbers is more favorable for antiproton collimation.

The first question we address is the particle removal rate. In the
experiment described in Figure~\ref{fig:removal}, the electron lens
was aligned and synchronized with the second antiproton bunch train,
and then turned on and off several times at the end of a collider
store. The electron beam current was about 0.4~A and the radius of the
hole was varied between 6$\sigma_y$ and 3.5$\sigma_y$, $\sigma_y =
\q{0.57}{mm}$ being the vertical r.m.s.\ beam size. The light-gray
trace is the electron-lens current. To isolate the effect of the
hollow beam, the ratio $\irel \equiv \Na/\Nc$ between the intensity of
the affected train~$\Na$ and the average intensity~$\Nc$ of the other
two control trains is shown in Figure~\ref{fig:removal} (black
trace). One can clearly see the smooth scraping effect. The
corresponding average removal rates $\Ndot = d\irel / dt$ are
collected in Table~\ref{tab:rates}.

\begin{figure}
\centering
\includegraphics[width=\figwid]{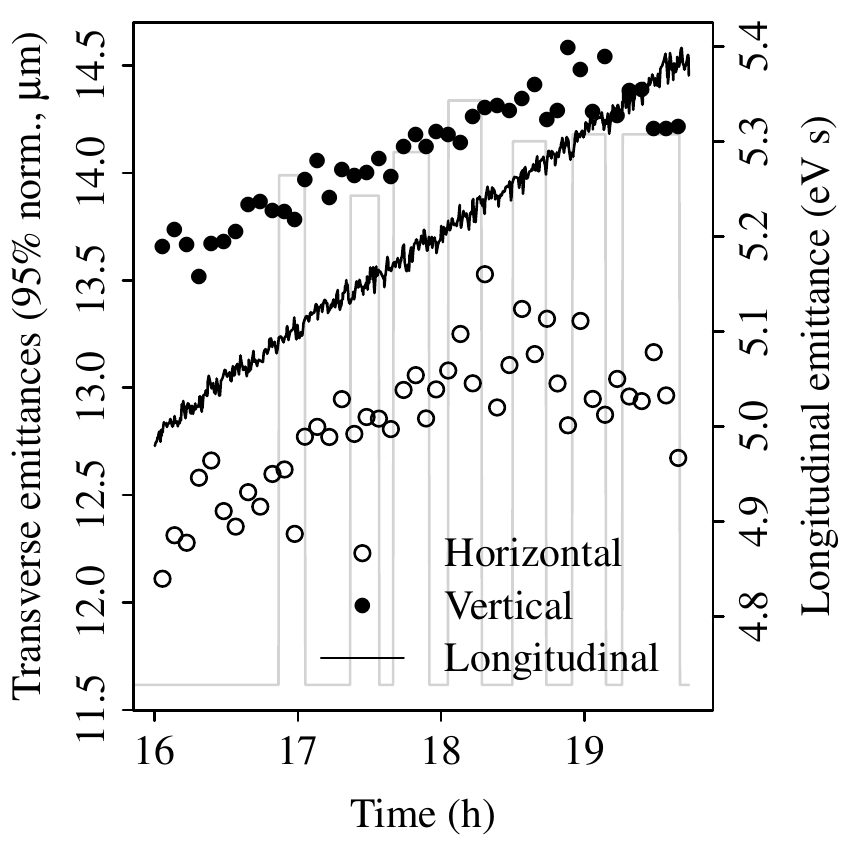}
\caption{Emittance evolution of the affected bunch train. The
  light-gray trace is the electron beam current (same experiment as
  Figure~\ref{fig:removal}).}
\label{fig:emit}
\end{figure}

Whether there are any adverse effects on the core of the circulating
beam is a concern, because the overlap region is not a perfect hollow
cylinder, due to asymmetries in gun emission, to evolution under space
charge of the hollow profile, and to the bends in the transport
system. We approached the problem from four points of view. First, one
can see from Figure~\ref{fig:removal} and Table~\ref{tab:rates} that
no decrease in intensity was observed with large hole sizes, when the
hollow beam was shadowed by the primary collimators. This implies that
the circulating beam was not significantly affected by the hollow
electron beam surrounding it, and that the effect on beam intensity of
residual fields near the axis was negligible.

\begin{figure}
\centering
\includegraphics[width=\figwid]{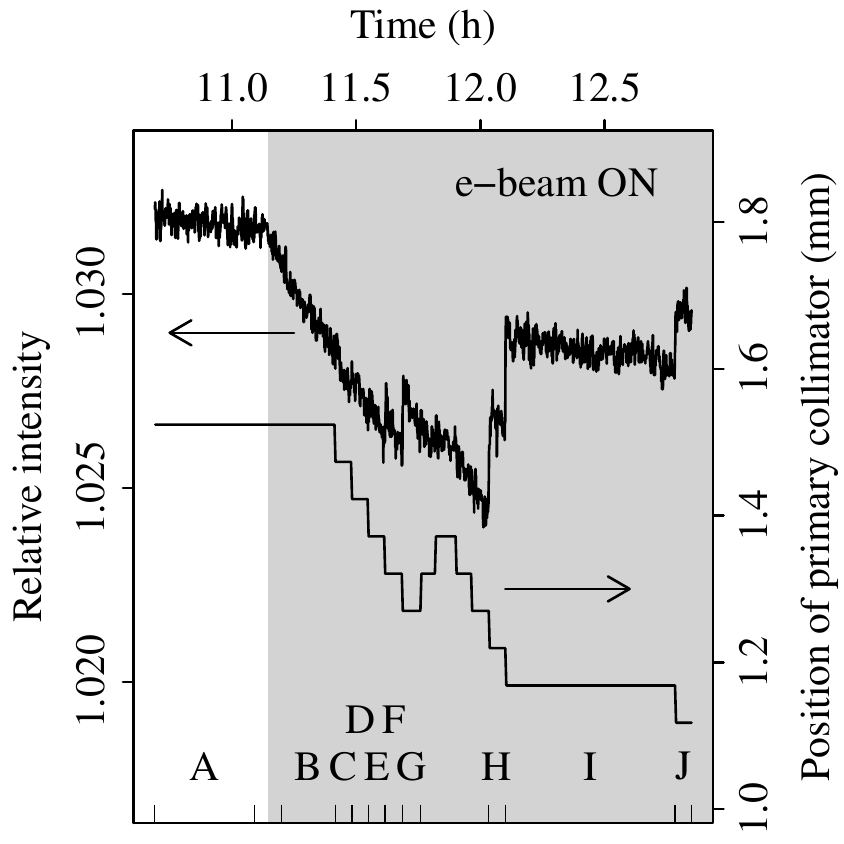}\\
\includegraphics[width=\figwid]{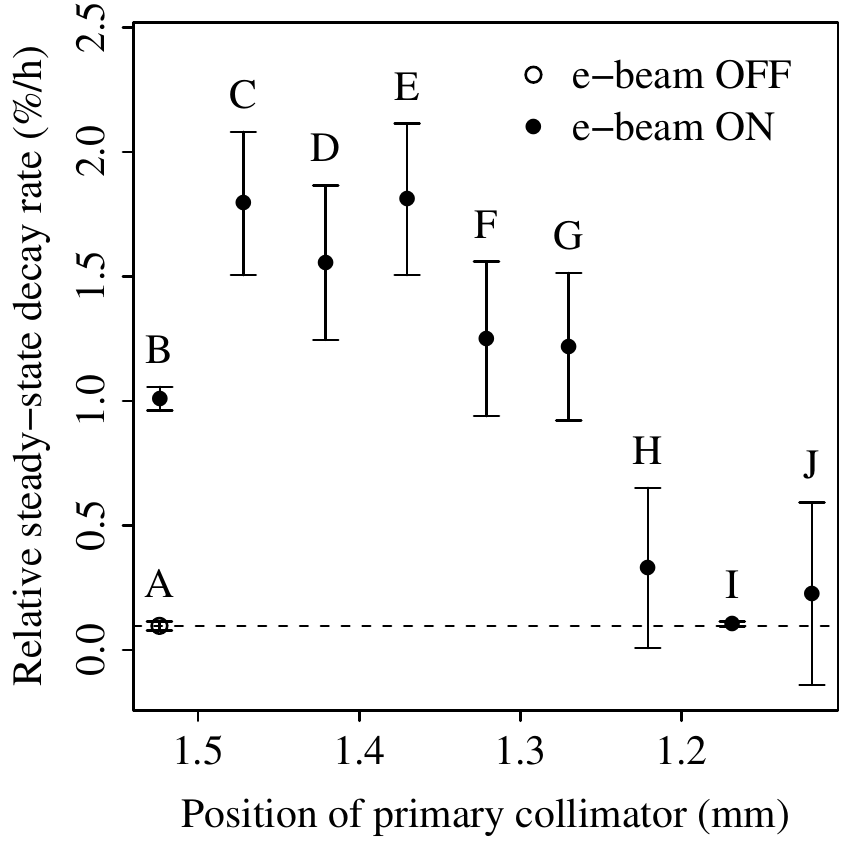}
\caption{Results of a collimator scan: (top)~relative intensity~\irel\
  of the affected bunch train and collimator distance from the beam
  axis vs.\ time; (bottom)~relative steady-state decay rate~\Ndot\ of
  the affected bunch train vs. collimator position, for each data set
  (letters A through J).}
\label{fig:coll}
\end{figure}

Secondly, one can observe the evolution of the emittances.
Figure~\ref{fig:emit} shows the average emittances of the affected
bunch train during the experiment of Figure~\ref{fig:removal}. If
there was emittance growth produced by the electron beam, it was much
smaller than that driven by the other two main factors, namely
intrabeam scattering and beam-beam interactions. As expected, for
small hole sizes, suppression of the beam tails translated into a
reduction in measured transverse emittances.

The effect of halo removal can also be observed by comparing beam
scraping with the corresponding decrease in luminosity. Luminosity is
proportional to the product of antiproton and proton populations, and
inversely proportional to the overlap area. If antiprotons are removed
uniformly and the other factors are unchanged, luminosity should
decrease by the same relative amount. If the hollow beam causes
emittance growth or proton loss, luminosity should decrease even
more. A smaller relative change in luminosity is a clear indication
that halo scraping is larger than core removal. In
Figure~\ref{fig:removal}, one can see how the luminosity for the
affected bunch~\La\ changed with time relative to the average
luminosity~\Lc\ of the control bunch trains. The gray trace is the
ratio $\lrel \equiv \La/\Lc$. The corresponding relative luminosity
decay rates $\Ldot = d\lrel / dt$ are reported in
Table~\ref{tab:rates}. The ratio between luminosity decay rates and
intensity decay rates increased with decreasing hole size.

Finally, one can attempt to directly measure the particle removal rate
as a function of amplitude. This was done with a collimator scan
(Figure~\ref{fig:coll}, top). A primary antiproton collimator was
moved vertically in 50-micron steps towards the beam axis. All other
collimators were retracted. The corresponding beam losses and decay
rates were recorded. The electron lens was acting on the second bunch
train with a peak current of 0.15~A and a hole size of 3.5$\sigma_y$,
or 1.3~mm at the location of the collimator. The corresponding
relative intensity decay rates~\Ndot\ as a function of collimator
position are shown in the bottom plot of Figure~\ref{fig:coll}. The
effect of the electron lens for a given collimator position is
represented by the difference between the~A and~B data sets. Data
sets~B through~J correspond to different collimator positions, all
with electron lens on. Particles are removed where electrons are, but
as soon as the primary collimator shadows the electron beam,
eliminating the halo at those amplitudes, the relative intensity decay
rate of the affected bunch train goes back to the value it had with
lens off. Even with such a small hole size, the effects of residual
fields on the core appear to be negligible. The time evolution of
losses during a collimator scan can also be used to measure changes in
diffusion rate as a function of amplitude~\cite{Seidel:1994}.

Losses generated by the electron lens were mostly deposited in the
collimators, with small changes at the experiments.  Alignment of the
beams was crucial, and the procedures based on the electron-lens
beam-position monitors were found to be reliable in spite of the
different time structure of the electron and (anti)proton pulses.  No
instabilities or emittance growth were observed over the course of
several hours at nominal antiproton intensities and electron beam
currents up to 1~A in confining fields above 1~T in the main
solenoid. Most of the studies were done parasitically during regular
collider stores.

In summary, it was demonstrated that controlled particle removal in
high-intensity storage rings and colliders with hollow electron beams
is viable. The device complements and extends conventional collimation
systems: particle removal is gradual and controllable, and the
electron beam can be placed arbitrarily close to the circulating
beam. To make the device more versatile, larger cathodes and higher
electron beam currents appear to be feasible, and experimental tests
in this direction are planned. Applicability to the Large Hadron
Collider is also under study.

\begin{acknowledgments}
  The authors would like to thank R.~Assmann and the CERN LHC
  Collimation Group, A.~Drozhdin, N.~Mokhov, and R.~Moore of Fermilab,
  and V.~Kamerdzhiev (Forschungszentrum J\"ulich, Germany) for
  discussions and insights; G.~Saewert (Fermilab) for the design of
  the high-voltage modulator; M.~Convery, C.~Gattuso, and T.~Johnson
  (Fermilab) for support during operation of the accelerator.

  Fermilab is operated by Fermi Research Alliance, LLC under Contract
  DE-AC02-07\-CH\-11\-359 with the United States Department of
  Energy. This work was partially supported by the U.S.\ LHC
  Accelerator Research Program (LARP).
\end{acknowledgments}

\end{document}